\documentstyle[11pt,newpasp,twoside]{article}
\def\edcomment#1{\iffalse\marginpar{\raggedright\sl#1\/}\else\relax\fi}
\reversemarginpar

\begin{document}
\title{VLA and MERLIN monitoring observations of the gravitational lens system B1030+074}
 \author{E. Xanthopoulos}
\affil{Jodrell Bank Observatory, University of Manchester, Macclesfield, Cheshire, SK11 9DL, UK}
\author{I. W. A. Browne}
\affil{Jodrell Bank Observatory, University of Manchester, Macclesfield, Cheshire, SK11 9DL, UK}
\author{A. R. Patnaik}
\affil{Max-Planck-Institut f\"{u}r Radioastronomie, Auf dem H\"{u}gel 69, D 53121, Bonn, Germany}
\author{P. N. Wilkinson}
\affil{Jodrell Bank Observatory, University of Manchester, Macclesfield, Cheshire, SK11 9DL, UK}
\begin{abstract}
We present VLA and MERLIN monitoring data of the JVAS  gravitational lens
system B1030+074. The system was monitored with the VLA from February 1998 to 
October 1998 at 8.4-GHz during which the VLA was at its A, BnA and B configuration. 
The 47 epochs of observations have an average spacing of approximately 5 days. 
Ten MERLIN snapshots were obtained in the  
L-band (1.7 GHz)  
during the months of April, May and June 
1998. 
Preliminary light curves of the two components of the lens system obtained from the VLA data 
indicate that during the period of the monitoring the A flux density showed a steady decrease. 
No changes are observed in the B light curve. 
\end{abstract}

\section{Introduction}

B1030+074 was discovered during the course of the Jodrell-Bank VLA Astrometric Survey 
(JVAS; Patnaik et al. 1992; Browne et al. 1998, Wilkinson et al. 1998)
which is a survey of flat-spectrum radio sources with the main purpose to search for 
gravitational lens systems. B1030+074 consists of two components separated by 1.56 arcseconds
with a flux ratio that seems to be changing with both time and frequency (Xanthopoulos et al. 1998).
Keck spectra have revealed a redshift of 0.599 for the lensing galaxy and of 1.535 for 
the background source. An initial model of this system has yielded a time delay of 156/h$_{50}$ days while
radio observations (EVN, MERLIN, VLBA, VLA) give a flux density ratio for the two 
components that ranges from 12.0 to 18.8. 
So since we already know the redshifts of the system and we are in the processs of improving the mass model 
of the lens (through VLBA observations; Xanthopoulos et al. 2000, in preparation) and also since the 
system has already shown signs of variability we embarqued on two monitoring programs 
with the main purpose to confirm the variable nature of the source.

\begin{figure*}
\begin{center}
\setlength{\unitlength}{1cm}
\begin{picture}(6,7)
\put(-1.8,-2){\includegraphics{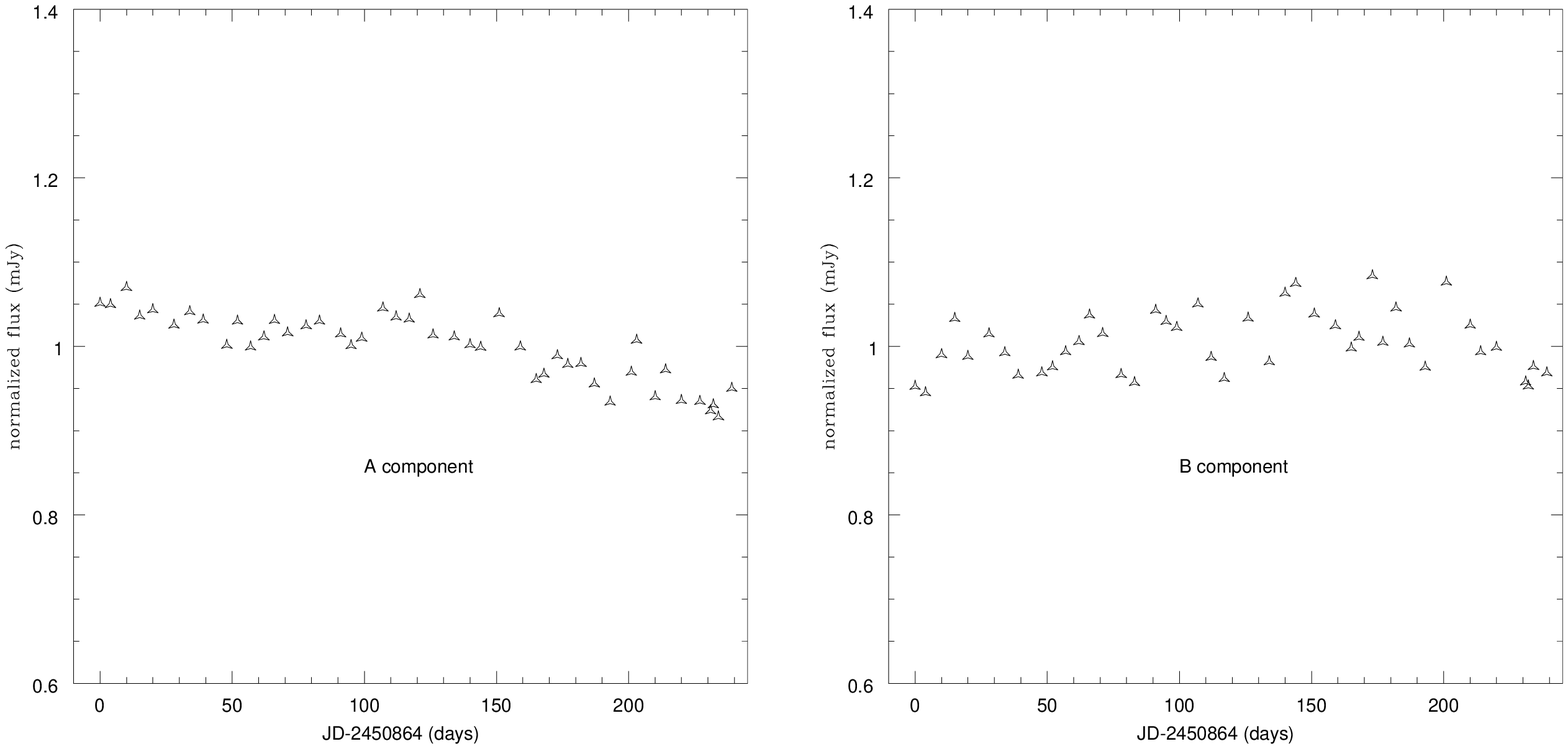}}
\end{picture}
\caption{Preliminary light curves of components A and B of the lens system B1030+074. }
\label{light}
\end{center}
\end{figure*}

\section{The Data and Preliminary Results}
The initial flux and phase calibration of the 47 epochs is done with {\sc aips} while the mapping 
and modelfitting was performed within {\sc difmap}. 
From this we obtained the light curves for components A and B of the lens system. 

Fig.~1 shows the preliminary light curves of B1030+074. 
We have divided each light curve by the average flux density of each component.
The light curve of component A shows a steady decrease in flux density with epoch
while we see no changes in the B light curve. Further analysis and interpretation 
of the results is underway.  
\section{References}
Browne, I. W. A., Patnaik, A. R., Wilkinson, P. N., Wrobel, J. B., 1998, MNRAS, 293, 257 

\noindent
Patnaik, A. R., Browne, I. W. A., Wilkinson, P. N., Wrobel, J. M., 1992, MNRAS, 254, 655

\noindent
Wilkinson, P. N., Browne, I. W. A., Patnaik, A. R., Wrobel, J. M., Sorathia, B., 1998, MNRAS, 300, 
790

\noindent
Xanthopoulos, E., et al. 1998, MNRAS, 300, 649. 

\end{document}